\documentclass[%
 reprint,
superscriptaddress,
 amsmath,amssymb,
 aps,
]{revtex4-2}

\usepackage{algorithm}
\usepackage{algorithmic}

\usepackage[utf8]{inputenc}
\usepackage{braket}

\usepackage{graphicx}
\usepackage{dcolumn}
\usepackage{bm}
\usepackage{hyperref}

\usepackage{mathrsfs}
\usepackage{todonotes}

\usepackage{amsthm}
\newtheorem{theorem}{Theorem}[section]

\newtheorem{proposition}[theorem]{Proposition}

\theoremstyle{remark}
\newtheorem{definition}[theorem]{Definition}

\usepackage{balance}

\begin{document}

\title{Untangling Surface Codes: Bridging Braids and Lattice Surgery}

\author{Alexandru Paler}
\email{alexandru.paler@aalto.fi}
\affiliation{Aalto University, 02150 Espoo, Finland}

\begin{abstract}
We present a systematic method for translating fault-tolerant quantum circuits between their braiding and lattice surgery (LS) representations within the surface code. Our approach employs the ZX calculus to establish an equivalence between these two paradigms, enabling verified, bidirectional conversion of arbitrary surface-code-level circuits. We show that both braiding and LS operations can be uniformly expressed as compositions of multibody measurements and demonstrate that the Raussendorf compression rule encompasses all known braid and bridge optimizations. We also introduce a novel CNOT circuit with LS. Our framework provides a foundation for the automated verification, compilation, and benchmarking of large-scale surface code computations, advancing toward a unified formal language for topological quantum computation.
\end{abstract}

\maketitle

\section{Introduction}

Fault-tolerant quantum computation relies on error-correcting codes to protect logical operations from noise. Among these, the surface code is the most practical and widely studied architecture for large-scale quantum computers. Logical operations in the surface code can be implemented using two distinct paradigms: \emph{braiding}, where topological defects are moved around each other, and \emph{lattice surgery} (LS), where logical qubit patches are merged and split through multibody measurements.

Although both paradigms are known to be computationally equivalent, they differ significantly in geometry, scheduling, and resource cost. Existing comparisons between them have been limited to specific circuits, and no general method exists for automatically translating or verifying equivalence between arbitrary braided and LS implementations. This lack of a formal translation layer hinders automated benchmarking, compiler validation, and optimisation across topological architectures.

In this work we present the first systematic framework for translating between braided and LS representations. Our approach leverages the ZX calculus~\cite{van2020zx, bombin2024unifying} to capture both paradigms within a unified, circuit-agnostic semantics. The resulting translation rules allow high-level surface-code circuits to be converted, verified, and
optimised automatically.

Beyond practical compiler applications, the framework also establishes a foundation for a formal language of surface-code computation, connecting ZX-calculus semantics with topological models of fault tolerance. A fully categorical treatment of these equivalences is deferred to future work, in which the underlying algebraic structures will be developed in detail.

\subsection{Contributions}


This work introduces a formally validated framework for translating fault-tolerant quantum computations between their \emph{braided} and \emph{lattice surgery} (LS) surface-code representations. The framework is based on circuit identities and the ZX calculus, which provides a semantics-preserving bridge between the two paradigms. To the best of our knowledge, this is the first systematic and bidirectional translation between arbitrary braided, LS, and quantum circuit descriptions.

\begin{enumerate}
    \item We define explicit translation maps between braided and LS circuits and show that they are semantics-preserving.
    \item We establish a systematic procedure for extracting high-level quantum circuits from arbitrary braided geometries, enabling automatic verification and benchmarking.
    \item We show that the Raussendorf compression rule subsumes all known braid and bridge optimisation techniques, including the bridging operation originally introduced in~\cite{fowler2012bridge}.
    \item We prove that bridges can be introduced and removed freely without altering the logical semantics, thereby enabling new classes of circuit transformations and simplifications.
    \item We present a new LS circuit for implementing the logical CNOT and prove its equivalence, via ZX-calculus reasoning, to their braided counterparts.
\end{enumerate}

While the constructions presented here can be expressed in a fully formal mathematical framework -- e.g., by defining explicit categories for braided, lattice-surgery, and ZX representations, and proving functorial equivalences between them—we have intentionally opted for an intuitive exposition. A completely formal treatment, while possible, would considerably increase the technical density and reduce accessibility for readers primarily interested in the computational implications. We therefore defer a rigorous categorical and proof-theoretic development of these translations to a follow-up paper, which will include a detailed formalisation and correctness proofs.

\subsection{Background}

We will introduce the minimum definitions necessary to present the translation method between braids, LS and circuits. We will consider that arbitrary quantum circuits have been compiled into the ICM form~\cite{paler2017fault, vijayan2024compilation}, which is built exclusively of: a) CNOTs, b) high fidelity non-Clifford states encoded in ancilla logical qubits, and c) logical qubit measurements. ICM circuits are fundamentally the way in which surface code computations are implemented. The lowest level surface code operations can be used only for CNOT gates, while universality is achieved through magic state ancilla and measurements.

Braided and LS circuits are executed in a measurement-based manner, and there is a Pauli frame that has to be tracked at the circuit's logical layer. The work of~\cite{de2020zx} has a complete treatment of how to track the corrections generated by the LS operations. In the following, we will not discuss the Pauli frame or how classically controlled gates have to be introduced into the quantum circuits or the ZX diagrams. We refer the reader to~\cite{de2020zx, paler2014software} for how to track and update the Pauli frame.

\subsubsection{Lattice Surgery}

In LS the two-dimensional lattice of physical qubits is partitioned into patches. Each patch has four boundaries, two for each operator type: logical X (bit) and logical Z (phase).

\begin{figure}[!h]
    \centering
    \includegraphics[width=0.8\columnwidth]{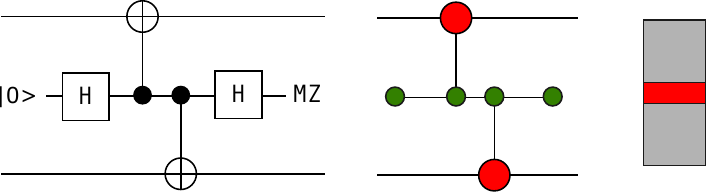}
    \includegraphics[width=0.8\columnwidth]{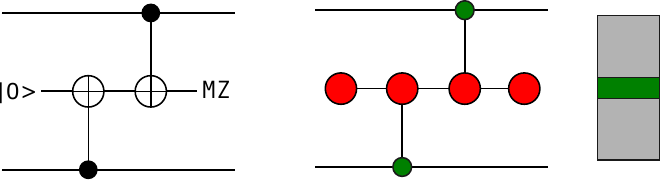}
    \caption{The two-body measurement circuits used throughout this manuscript. The XX-measurement (top) uses an ancilla qubit initialized in $\ket{+}$ and measured in the X-basis and has an equivalent ZX-diagram where two red spiders are touching the measured wires. The XX-measurement can be abstracted in the language of lattice surgery (LS) as two grey patch interacting along their red boundaries. The ZZ-measurement (bottom) uses an ancilla initialized in $\ket{0}$ and measured in the Z-basis. The LS diagram uses two grey patches interacted along their green boundary.}
    \label{fig:zzxx}
\end{figure}
\begin{figure}[!h]
    \centering
    \includegraphics[width=0.95\columnwidth]{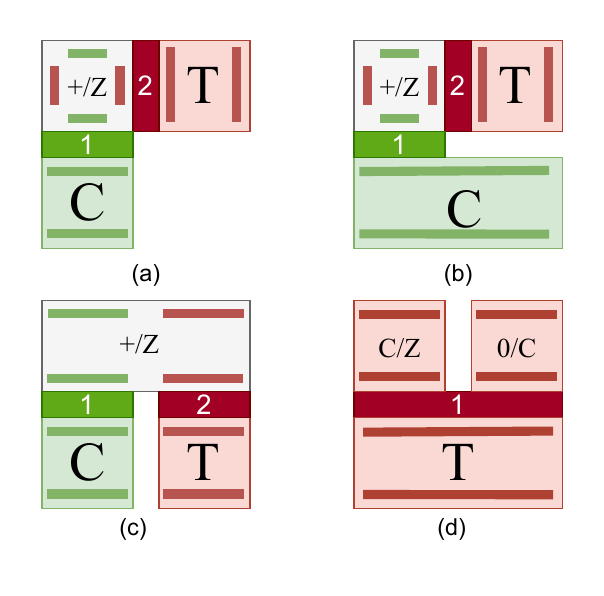}
    \caption{There are different ways of implementing CNOTs with surface code lattice surgery. In the following the color of the square patches is indicative of the patch boundary that is about to be used. The $+/Z$ notation means that the patch will be initialized in $\ket{+}$ and measured in the Z-basis. The indices on the two-body measurements (cf. Fig.~\ref{fig:zzxx}) indicates their order. For example, in (a) the green ZZ-measurements is performed first, and the red XX-measurement is executed afterwards. (a) The canonical patch arrangement for LS CNOT~\cite{horsman2012surface}; (b) extending one of the patches (e.g. the control $C$) does not change the computation; (c) after extending and rotating the ancilla patch such that it has two operator boundaries on the same side~\cite{litinski2019game};  (d) the ancilla is now initialised in $\ket{0}$ and not measured, the control patch is measured in the Z basis after all three patches performed a three-body red X-measurement. The correctness of (d) is shown using ZX calculus in Fig.~\ref{fig:correctlspatchcnots}.}
    \label{fig:lspatchcnots}
\end{figure}

In LS, patches are operated by merging and splitting them along the boundaries. The merge and split operations correspond to multibody measurement circuits (Fig.~\ref{fig:zzxx}). The same operations are also used for implementing the CNOT gate (Fig.~\ref{fig:lspatchcnots}). Multibody measurements are operations in which multiple patches are merged and split simultaneously. The connection between lattice surgery and the ZX calculus has been formalised by~\cite{de2020zx}, and one of the earliest works that has used this connection for expressing lattice surgery compilation tasks was~\cite{gidney2019flexible}.

The first CNOT implementation was presented in~\cite{horsman2012surface}, but, as shown in~\cite{de2020zx}, there are more possible constructions. Computational universality with LS is achieved by injection and distillation procedures which are mapped to logical qubit patches and CNOT gates implemented through merge and split operations.

\subsubsection{Braids}

Computations implemented by braiding are achieved after punching holes (also called defects) into the physical qubit lattice. The braids are the result of performing the following abstractions: a) filling out the holes with a solid; b) representing the time evolution (resulting from the discretised movement across the lattice) of the holes by strands; c) removing everything around the strands, including the lattice.

In the canonical compilation of braided circuits~\cite{paler2014design}, logical qubits are defined by pairs of strands. There are two types of strands: primal and dual. Accordingly, depending on the strand types, there are primal and dual logical qubits.

\begin{figure}
    \centering
    \includegraphics[width=0.9\columnwidth]{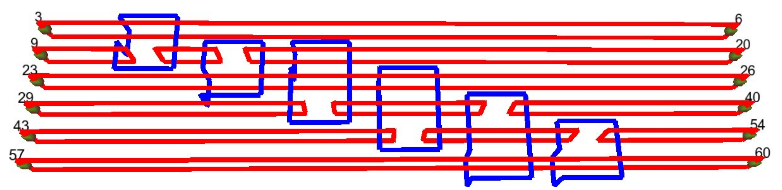}
    \caption{Elements of braided circuits~\cite{paler2017fault}: pairs of defects (here primal, red) form logical qubits and are braided with loops of the opposite type (here dual, blue) for performing CNOTs. The circuit herein does not include bridges.}
    \label{fig:braidprimalcnot}
\end{figure}

\begin{figure}[!h]
    \centering
    \includegraphics[width=0.9\columnwidth]{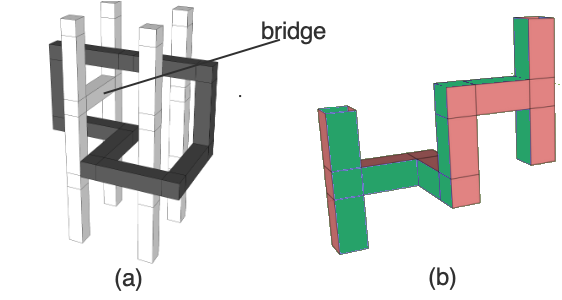}
    \caption{CNOT implementations: (a) using braided surface codes~\cite{fowler2012bridge} between two primary logical qubits (here white formed by pairs of defects) mediated by a dual loop (here dark grey); (b) lattice surgery. Time flows in both diagrams from the bottom to the top. The diagrams are visually extremely similar. However, this is just the result of the white \emph{bridge} that is connecting the pair of vertically parallel defects in (a). In (b), we used green and red and to highlight the different correlation surfaces spanned along the patches.}
    \label{fig:cnots}
\end{figure}

Strands can be either \emph{braided} (Figures~\ref{fig:braidprimalcnot} and~\ref{fig:cnots}) or \emph{bridged} (Fig.~\ref{fig:cnots}a) . Braiding works for strands of the opposite type (primal-dual), while bridging is available for strands of the same type (primal-primal, dual-dual).

\subsubsection{Correlation Surfaces}

Correlation surfaces have been initially proposed in the cluster-state implementation of the surface code~\cite{raussendorf2007topological}. A first method for automatically constructing correlation surfaces has been proposed in~\cite{paler2012synthesis}, and then a construction using Boolean variables in~\cite{paler2014design}. A SAT based approach to constructing LS correlation surfaces has been described by~\cite{tan2024sat}.

\begin{definition}
A correlation surface as the set of qubits tracked along the path connecting the input logical operator with the output logical operator of a surface code computation.
\end{definition}

In the case of braided computations, there are two types of correlation surfaces: tubes and sheets. A \emph{tube} is formed by the logical operators encircling the strands. A \emph{sheet} is formed by the logical operators spanned between pairs of strands (e.g. in a logical qubit formed by a pair of strands). The interpretation of the tubes and sheets, in terms of logical X and logical Z operators, alternates between the types of logical qubits (primal or dual).

\begin{figure}[!h]
    \centering
    \includegraphics[width=\columnwidth]{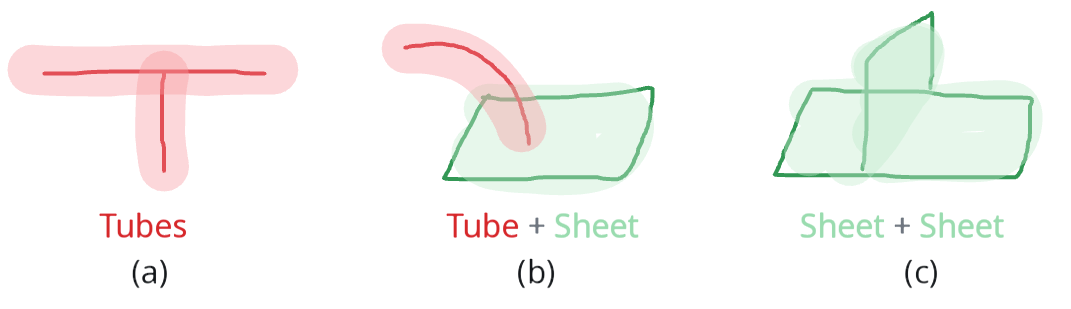}
    \caption{Connecting correlation surfaces. Here we show how primal-red can be connected to dual-green. The reverse are also valid. a) two tubes of the same type; b) a tube and a sheet of opposite type; c) two sheets of the same type. Herein, the red strings correspond to the white defects from and the green loops to the grey defect loop from Fig.~\ref{fig:cnots}.}
    \label{fig:poss}
\end{figure}

\begin{figure}[!h]
    \centering
    \includegraphics[width=0.8\columnwidth]{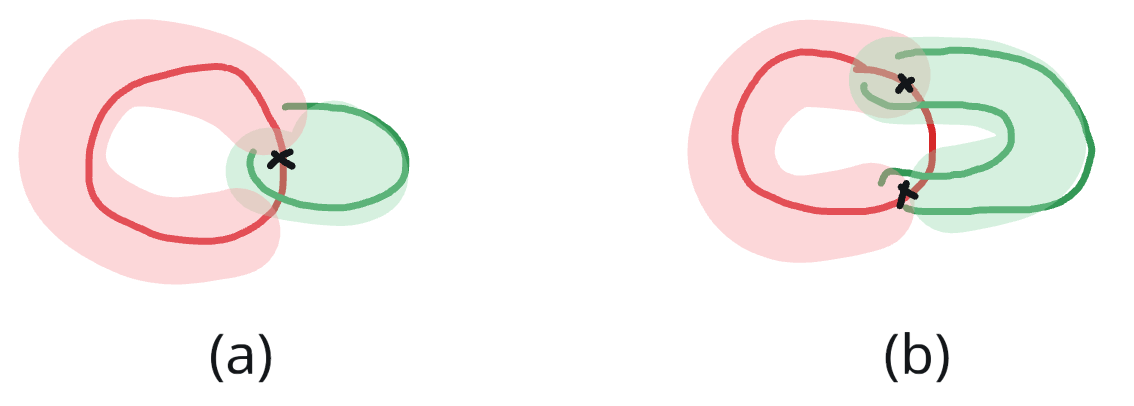}
    \caption{Two loops of opposite types have to be braided more than once, because otherwise the correlation surfaces would not be correctly connected with each other. This observation is a criteria for checking the correctness of braided circuits. (a) incorrect connection of a red tube to the green sheet, because the tube would need to connect twice at the black braiding point; (b) the red tube can connect correctly to the green sheet, because now there two braiding points and each of these points is touched only once by the tube.}
    \label{fig:possc}
\end{figure}

Figs.~\ref{fig:poss} and\ref{fig:possc} illustrate the correct ways in which tubes and sheets can be connected: a) tubes connect to a sheet, when the supporting strands are of opposite type; b) tubes connect to tubes, or sheets to sheets, when the supporting strands are of the same type.

The construction from Fig.~\ref{fig:poss}a exists wherever two strands are \emph{bridged}~\cite{fowler2012bridge}: the tubes have to be connected in order for the correlation surface to be well-defined. At the same time, the sheets spanned along the bridged strands will be overlapping (Fig.~\ref{fig:poss}c) such that there are three valid constructions. The construction from Fig.~\ref{fig:poss}b corresponds to a braid: tubes will be connected to sheets. Bridging and unbridging will be discussed in Sec.~\ref{sec:unbridge}.

\begin{definition}
In the case of lattice surgery (LS), we define correlation surfaces as the sheets covering the boundaries of the patches (when one considers these in 3D in space-time diagram-like manner like in Fig.~\ref{fig:cnots}). 
\end{definition}

In LS, we do not consider a dual lattice space, such that there are no primal or dual logical qubits. For this reason, the sheet correlation surfaces will correspond to logical X and logical Z operators at the boundaries of the patches. Due to the way how logical operators are initialised in braided computations, well-defined sheets exist only in loops, and well-defined tubes only in non-loops (i.e. chains).

\subsubsection{Trivial Loops}

\begin{figure}[!h]
    \centering
    \includegraphics[width=\columnwidth]{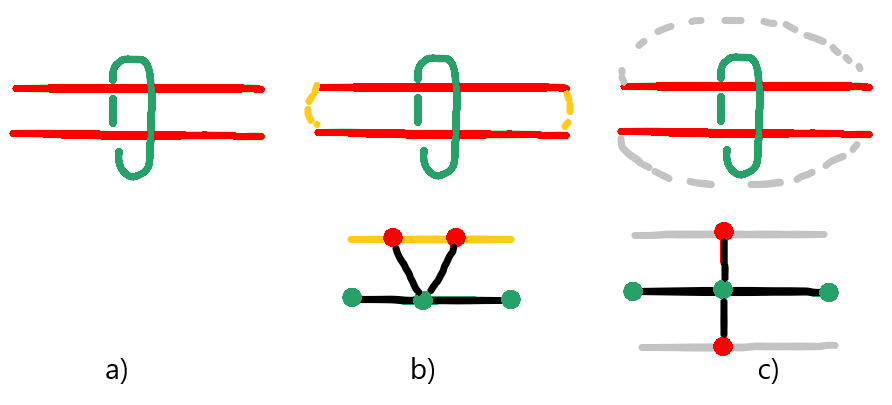}
    \caption{a) A loop around a pair of opposite type strands; b) if the pair belongs to the same loop then it represents an identity as shown by the ZX diagram; c) if the strands belong to different loops, the result is a multibody measurement, in this example an XX.}
    \label{fig:ring1}
\end{figure}

Our focus are loops, and we start by discussing a single loop surrounding a pair of strands (Fig.~\ref{fig:ring1}). The blue loop in Fig.~\ref{fig:ring1}b is just an identity from a computational point of view, but as shown in the Appendix it cannot be simply removed. That ring is an identity only if the two strands belong to the same logical qubit, otherwise, if the strands are from distinct qubits, the ring is a parity measurement.

\begin{definition}
    A loop is trivial when it acts like an identity operator (Fig.~\ref{fig:ring1}b).
\end{definition}

\begin{definition}
    A loop is not trivial when it acts like an multibody operator (Fig.~\ref{fig:ring1}c).
\end{definition}

Trivial loops are a necessity when optimising braided structures, because it allows eliminating other loops and strands (Sec.~\ref{sec:raussendorf}). We can introduce trivial loops using the ZX calculus (Fig.~\ref{fig:ring2}) and can decompose multibody measurements into pairs of non-trivial rings such as the one from Fig.~\ref{fig:ring1}c. 

A ring around a loop of an opposite type (Fig.~\ref{fig:ring1}c) is a ZZ- or XX-measurement, depending on the type of the loop in the middle. The sheet of the outer loop can be used to measure the sign of the logical operators that are correlated along both tubes. This is effectively, in terms of LS, a multibody measurement between a qubit and an ancilla.

\section{Methods}

We offer a new perspective on LS and braided circuits, and will explain why the resemblance between bridged braided circuits and LS is more than a coincidence. We use the ZZ and XX measurement circuits (parity circuits, Fig.~\ref{fig:zzxx}) and the ZX calculus to extract universal quantum circuits in order to show the equivalence between braided and LS diagrams (Algorithm~\ref{alg:translation}). In the process of doing this, we introduce two reduced instructions sets: \texttt{LSIS} for LS circuits, and \texttt{RBIS} for braided circuits.

Our goal is to interpret braided and LS structures as ICM circuits. In an ICM circuit, the CNOT gate is used for achieving computational universality\cite{paler2017fault}: the T gates are performed by teleportation. We will not discuss the exact implementation of the T gates, or how non-Clifford states are encoded (i.e. injection) and their fidelity is increased (i.e. distillation), because those procedures use circuits formed exclusively of CNOT gates.

We use the CNOT gate to mediate the translation between LS and braids:
\begin{itemize}
    \item braids-LS direction: we use the ZX interpretation of the braided structures to extract LS multibody measurements;
    
    \item LS-braids direction: LS computations are expressed as sequences of multibody measurements, and we will express the measurements using the circuits from Fig.~\ref{fig:zzxx} and then translate those circuits into braids.
\end{itemize}

For both translation directions, LS-braids and braids-LS, we assume, for the beginning, that the braided structures are formed entirely of simple loops and strands. Afterwards, we will show how to decompose arbitrarily complex braided structures into simple loops and strands.

We will follow a bottom-up approach. 
\begin{itemize}
    \item first, we introduce the straightforward bidirectional translation between LS and braids without bridges;
    \item second, we analyze, from the perspective of correlation surfaces, how multibody measurements are implemented in braided and LS circuits;
    \item finaly, we present methods for the automatic translation of arbitrary braided diagrams into LS and vice versa.
\end{itemize}

\begin{figure}[t]
\begin{algorithmic}[1]
\REQUIRE A surface-code circuit $C$, represented either as a braided diagram $B$ or as a lattice-surgery (LS) diagram $L$
\ENSURE The corresponding circuit in the alternate representation
\IF{$C$ is a braided circuit $B$}
    \STATE Identify all \textbf{loops} and \textbf{bridges} in $B$
    \STATE Decompose $B$ into \emph{elementary loops} using unbridging rules
    \FORALL{braids between loops of opposite type (primal--dual)}
        \STATE Replace each braid with an \textbf{XX} or \textbf{ZZ} multibody measurement
        \STATE Express the measurement as LS merge/split operations on logical patches
    \ENDFOR
    \STATE Apply ZX simplifications to contract trivial spiders
    \STATE Return the LS circuit
\ELSE
    \STATE Identify all \textbf{merge/split} operations and their measurement bases in $L$
    \FORALL{multibody measurements}
        \STATE Express the measurement as a ZX diagram
        \STATE Translate each two-body ZX interaction into a braid between opposite-type strands
    \ENDFOR
    \STATE Introduce bridges for same-type strand interactions if needed
    \STATE Apply loop compression (Raussendorf rule) to reduce redundant braids
    \STATE Return the braided circuit
\ENDIF
\end{algorithmic}
\caption{Bidirectional Translation Between Braided and Lattice Surgery Circuits}
\label{alg:translation}
\end{figure}

\subsection{The braids-LS translation: Loops and Spiders}
\label{sec:bls}


The braid-LS translation begins by selecting the inputs and the outputs of the braided structure. In general, a braided circuit is the diagrammatic equivalent of an ICM circuit, and the order in which the CNOT gates are executed is determined by the topological sort from diagram inputs towards outputs.

Strands and loops are the building blocks of braided circuits. However, without loss of generality, in order to keep the LS-braids interpretation straightforward, we use only loops for representing logical qubits. This approach is slightly different from the original~\cite{raussendorf2007topological}, where pairs of strands were the support of a logical qubit: loops resulted depending on the initialisation and measurement basis.

For our purpose, logical qubits initialised into $\ket{0}$ will be primal loops, for $\ket{+}$ the loops will be duals. For these reasons: a) primal loops will correspond to logical qubits initialised and measured in the Z basis; b) dual loops correspond to logical qubits initialised and measured in the X basis.

There are similarities between loops and spiders. For example, it has been observed by~\cite{hanks2020effective} that the double strand approach to defining braided logical qubits is similar to constructing GHZ states. The above example highlights the applicability of the ZX calculus for braided circuit analysis~\cite{hanks2020effective}. This has been also very recently observed in~\cite{kupper2025string}, but that paper does not discuss bridges, because these are not effectively braids.

\begin{figure}[!t]
    \centering
    \includegraphics[width=0.8\columnwidth]{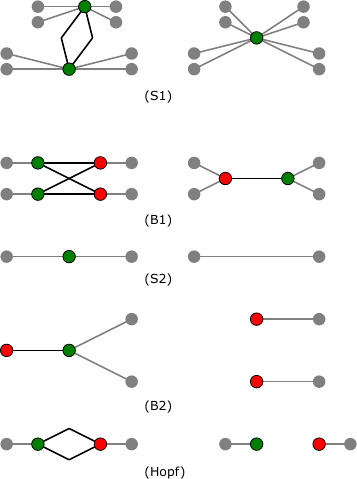}
    \caption{The transformation rules of the ZX calculus~\cite{van2020zx}. The rules are valid also when spiders are switching their color. For example, the (S2) rule says that non-phased spider can be removed or trivially added to a wire irrespective of its color (green or blue).}
    \label{fig:zx_rules}
\end{figure}

The main difference between ZX wires and strands is that the first do not have a type (colour), while the latter do. For this reason, it is useful to imagine that, for a spider with arity larger or equal to two, two of the spider's legs correspond to the qubit-wire in the circuit diagram, while the other legs are for CNOTs. This observation uses the fact that a ZX diagram is a bipartite graph: there are no adjacent same-colour spiders, and, even if such a situation would exist, the spiders can be merged (Fig.~\ref{fig:zx_rules}). In other words, a ZX spider can be translated into a braided loop, by: 1) having two of the spider's legs as a loop; 2) turning the remaining legs into  braids (CNOTs).

\subsection{The LS-braids translation: Rotations and Braids}
\label{sec:lsb}

\begin{figure}
    \centering
    \includegraphics[width=\columnwidth]{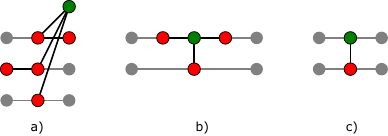}
    \caption{The correctness of Fig.~\ref{fig:lspatchcnots}d): a) the three patches are horizontal wires connected by the ZX diagram of a 3-body measurement; b) after contracting all red spiders; c) eliminating the trivial red spiders results in the ZX diagram of a CNOT.}
    \label{fig:correctlspatchcnots}
\end{figure}

The LS-braids translation is based on the observation that the multibody measurements for the LS CNOT are structurally (and functionally) similar to the braiding pattern of the primal-primal CNOT. Fig.~\ref{fig:correctlspatchcnots}d highlights the connection between the LS CNOT and the braid CNOT: both are the result of implementing the same three-body X measurement and then measuring the control qubit/patch while maintaining the qubit/patch of the ancilla initialised in the Z basis.

Similarly to how~\cite{litinski2019game} introduced a set of rules in order to express arbitrary LS computations, we define a \texttt{LS reduced instruction set (LSIS)} in order to translate between LS and braids:
\begin{itemize}
    \item \emph{initialisation into $\ket{0}$ or $\ket{+}$} and \emph{measurement into X or Z basis} (e.g. the $0/Z$ notation on the ancilla patches from Fig.~\ref{fig:poss});
    \item \emph{extend and contract} -- a patch is extended to the next empty patch space, or contracted to leave an empty patch space (e.g. Fig.~\ref{fig:lspatchcnots}b);
    \item \emph{rotate} -- the orientation of the boundaries is changed; for example, the X boundary is moved from the top to the right (e.g. the effect on the T-patch when comparing Fig.~\ref{fig:lspatchcnots}a and Fig.~\ref{fig:lspatchcnots}c);
    \item \emph{merge and split} -- necessary for implementing multi-body measurements (e.g. the operations from .
\end{itemize}

The LS patch rotation operation is switching the position of the logical operators.  The rotation~\cite{horsman2012surface, litinski2019game} is used to position logical operators such that the necessary multibody measurement can be implemented. In~\cite{fowler2018low}, a rotation was assumed to take twice the code distance, such that rotations had to be performed at half distance -- a procedure assumed to be \emph{sufficient} for practical purposes. In~\cite{litinski2019game} a rotation takes three times the code distance. Rotations could also be implemented with Hadamard gates, which in the case of the surface code are transversal~\cite{horsman2012surface}. This would reduce the depth of the compiled circuit. If sufficent hardware resources are available, it is not necessary to rotate a patch in order to access its logical operator boundary~\cite{litinski2019game, watkins2024high}.

When braiding, this LS rotation is equivalent to changing the type of the correlation surface corresponding to the logical operator: a tube becomes a sheet, and a sheet is turned into a tube (Fig.~\ref{fig:poss}). 

\begin{proposition}
A LS rotation corresponds to a braid between strands of different types.
\end{proposition}

\begin{figure}[!h]
    \centering
    \includegraphics[width=\columnwidth]{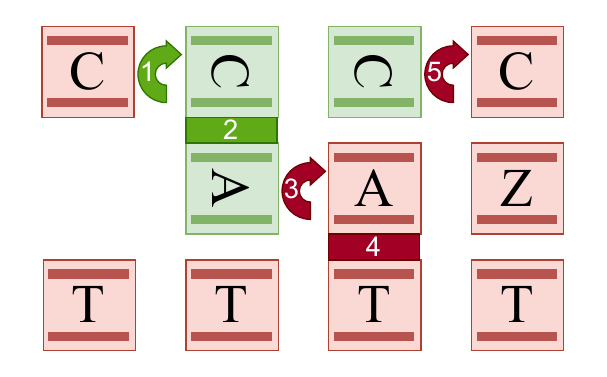}
    \includegraphics[width=\columnwidth]{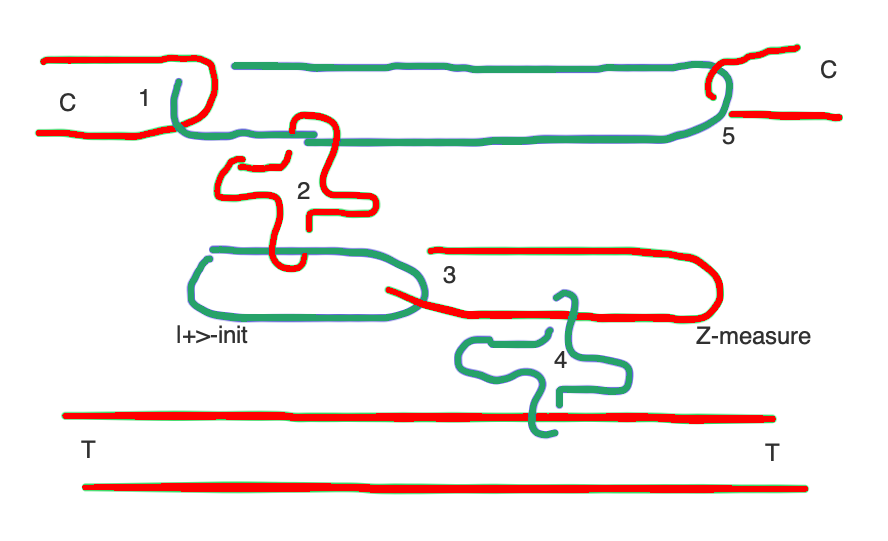}
    \caption{Parallel comparison between the LS CNOT and the braided CNOT using XX and ZZ measurements. The control and dual qubits are primals, the ancilla because it is initialised in $\ket{+}$ it is directly a dual. The rotation arrows correspond to braiding a primal with a dual. This one-to-one comparison between circuits is practically a visual translation of \texttt{LSIS} (Sec.~\ref{sec:lsb}) to \texttt{RBIS} (Sec.~\ref{sec:unbridge}).}
    \label{fig:rotations}
\end{figure}

We will now illustrate how Fig.~\ref{fig:lspatchcnots}a can be automatically translated into the standard functionally equivalent braided circuit~\cite{raussendorf2007topological}. We start from the LS CNOT circuit from Fig.~\ref{fig:lspatchcnots}a. The latter includes two multi-body measurements, XX and ZZ, which correspond to the circuits from Fig.~\ref{fig:zzxx}. The step-by-step constructive translation of a CNOT from LS to braids is illustrated in Fig.~\ref{fig:rotations}.

\subsection{Multibody Measurements and the Bialgebra Rule}

We translated a LS CNOT into braids, and used the ZX language for expressing two-body measurements. We will use two-body measurements to compose multi-body measurements compatible with LS and braids. The two-body measurements are effectively the key of our translation methods.

In braided circuits, the simplest multibody measurement patterns appear, for example, when a single loop is braided with multiple other loops of the opposite type. In LS, multibody measurements appear each time that merge and split operations are executed (e.g. Fig.~\ref{fig:lspatchcnots}). 

\begin{figure}[!h]
    \centering
    \includegraphics[width=\columnwidth]{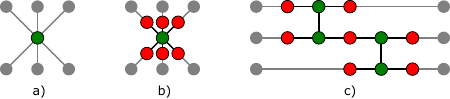}
    \caption{Manipulating spiders: a) a green spider connecting multiple wires; b) trivial red spiders are introduced; c) the green spider is split, another trivial red spider is included between the two greens, and the wires are arranged to look like a quantum circuit. }
    \label{fig:pairwise}
\end{figure}

\begin{proposition}
In the ZX calculus, spiders are manifestations of a multibody measurement~\cite{van2020zx}.
\end{proposition}

Starting from Fig.~\ref{fig:pairwise}a, we use the ZX calculus to split a spider into a sequence of two-body measurements illustrated in Fig.~\ref{fig:pairwise}c: the pairwise green spiders are two-body measurements which can be easily translated into braids if the red spiders are not trivially eliminated. If the red spiders are trivially removed, the green spiders can be easily translated into LS merge and split operations.

\begin{figure}[!h]
    \centering
    \includegraphics[width=\columnwidth]{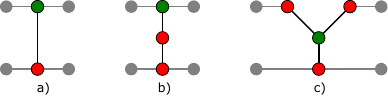}
    \includegraphics[width=\columnwidth]{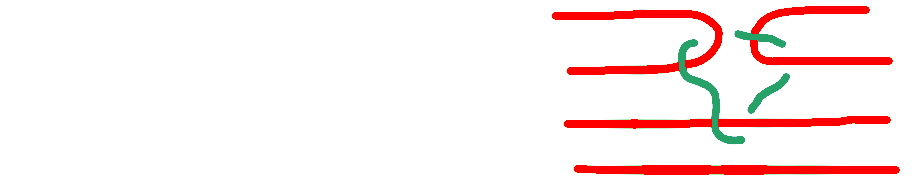}
    \caption{An example of how to use the bialgebra rule to build a multibody measurement mediated by a green loop. Effectively, this is a demonstration of how a CNOT is implemented in braided diagrams: a) a CNOT in the ZX calculus; b) after introducing a trivial red spider; c) after applying the bialgebra rule and translating to braids.}
    \label{fig:gencnot}
\end{figure}

In Fig.~\ref{fig:gencnot} we derive the braided primal-primal CNOT step-by-step by starting from the ZX diagram of a CNOT. We chose a seemingly more complex route to arrive at the multibody measurement implemented by the red spider: applied the bialgebra rule. Moreover, the bialgebra rule shows that it is possible to achieve a braided CNOT using four loops (Fig.~\ref{fig:genallbialgebra}).

We can generalise the construction of multibody measurements. From the perspective of LS/braids, these measurements appear by using an ancilla qubit (Fig.~\ref{fig:genallbialgebra}) initialised in a given basis, interacting it with patch-operators/strands of the opposite type, and measuring the ancilla in the opposite basis.

\begin{figure}[!h]
    \centering
    \includegraphics[width=\columnwidth]{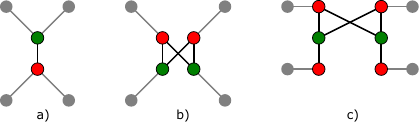}
    \includegraphics[width=\columnwidth]{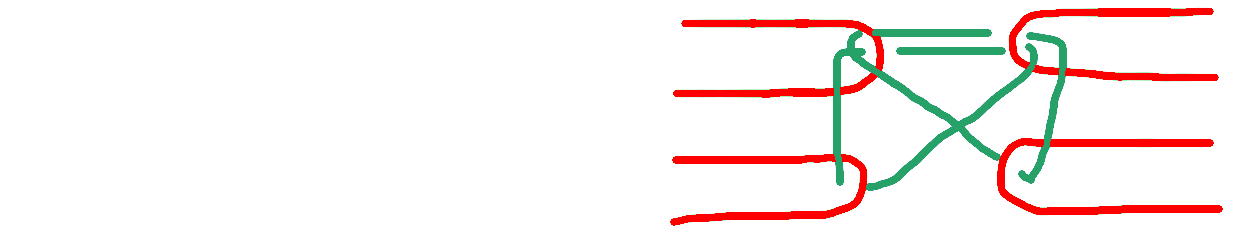}
    \caption{A slightly more complex application of the bialgebra rule that results in two 3-body measurements: a) a CNOT; b) applying the bialgebra rule; c) after rearranging the wires and translating into braids after inserting trivial green spiders on the lower wires.}
    \label{fig:genbialgebra}
\end{figure}

\begin{figure}[!h]
    \centering
    \includegraphics[width=\columnwidth]{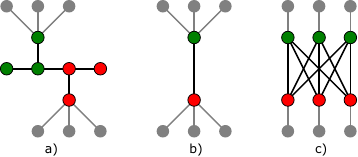}
    \caption{Two spiders of opposite colours can be interpreted as sharing an ancilla initialised and measured in opposite basis: a) the ancilla wire is highlighted; b) the general input to the bialgebra rule; c) after applying the bialgebra rule. The last subfigure can be transformed, for example, into braids in a similar way to Fig.~\ref{fig:genbialgebra}.}
    \label{fig:genallbialgebra}
\end{figure}

\subsection{Spiders and Bridges}

Multibody measurements can be analysed also by their effect on correlation surfaces. The latter are helpful to determine a method to decompose seemingly complex braided structures into elementary braided loops. In the translation from Sec.~\ref{sec:bls}, we used only loops in braided circuits. Due to this decision, when using ZX diagrams to interpret braids, the colour of the spiders will play a significant role with respect to where and how the correlation surfaces will be interacted.

\begin{proposition}
In braided circuits, every two-body measurement can be replaced with a bridge.
\end{proposition}

Assuming, without loss of generality, that all the wires of a ZX diagram are translated into braids as primal-red loops:
\begin{itemize}
    \item ZZ-measurements are used to interact sheets (logical Z operators);
    \item XX-measurements are used to interact tubes (logical X operators).
\end{itemize}

Although XX- and ZZ-measurements are used for different tasks, internally the circuits perform the same thing: they correlate sheets, the difference being only the space where this operation takes place (primal vs. dual).

\begin{proposition}
    Bridging does not change the computation and does not affect the functional correctness of the diagrams.
\end{proposition}

\begin{figure}[!h]
    \centering
    \includegraphics[width=\columnwidth]{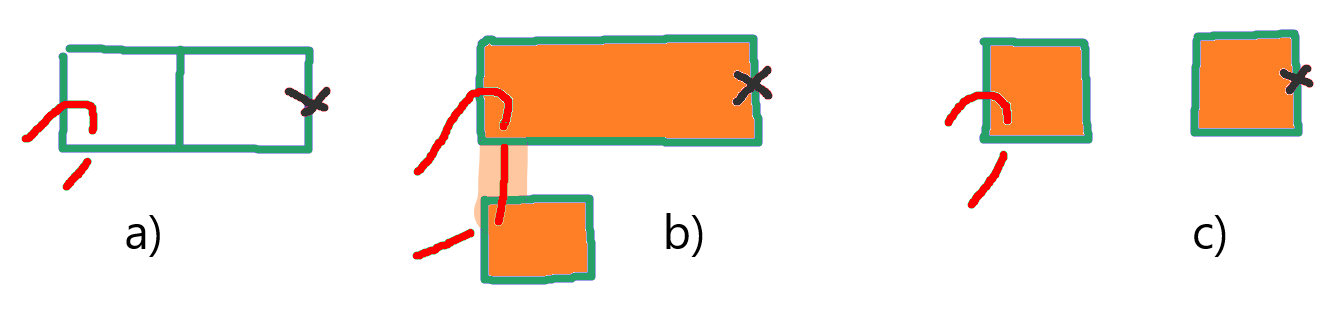}
    \caption{Example of loops after unbridging: a) the original 3D structure, b) disconnecting loops along braided defect (red); c)  disconnecting disjoint loops (i.e. no other defect is braiding them). The black cross marks that the associated loop may continue and be braided with another dual.}
    \label{fig:unbridge}
\end{figure}

From the perspective of the ZX calculus, the type of the spider determines the type of the bridge. Thus, bridges between strands of the same type/color can be arbitrarily introduced into the braided structure. We show this by case distinction:
\begin{enumerate}
    \item The bridge is between strands belonging to the same loop (Fig.~\ref{fig:unbridge}b). This is a special case in the step by step derivation of the Raussendorf rule: 1) only two primal loops are present; 2) no dual moustaches; 3) the central strand is not removed. None of the above change the computation after bridging;
    
    \item The bridge is between two distinct loops (Fig.~\ref{fig:unbridge}c). If there is no chain of braids connecting the loops then the entire structure consists of two disjoint sheets and their trivial product that results after the bridge product -- effectively there is no tube than enforces the sheet to be connected to it (cf. Figures~\ref{fig:poss} and \ref{fig:possc})
\end{enumerate}

\subsection{Unbridging: Decomposing Complex Braided Structures}
\label{sec:unbridge}

LS circuits are straightforward to translate into to braided circuits. Depending on the type of LS multibody measurement the corresponding braided structure can be automatically generated. Translating a braided circuit to LS does not seem as easy, because braided circuits might include bridges. We need independent, not bridged~\cite{fowler2012bridge}, loops (e.g. Fig.~\ref{fig:poss}a). If there are bridged loops in a diagram, the bridges representation have to be undone, meaning that each bridged structure is decomposed into elementary loops.

\begin{definition}
We call \emph{unbridging} the procedure by which a braided circuit is transformed such that all correlation surface combinations of the type Fig.~\ref{fig:poss}a and Fig.~\ref{fig:poss}c are transformed into the type from Fig.~\ref{fig:poss}b.
\end{definition}

\begin{figure}[!h]
    \centering
    \includegraphics[width=0.8\columnwidth]{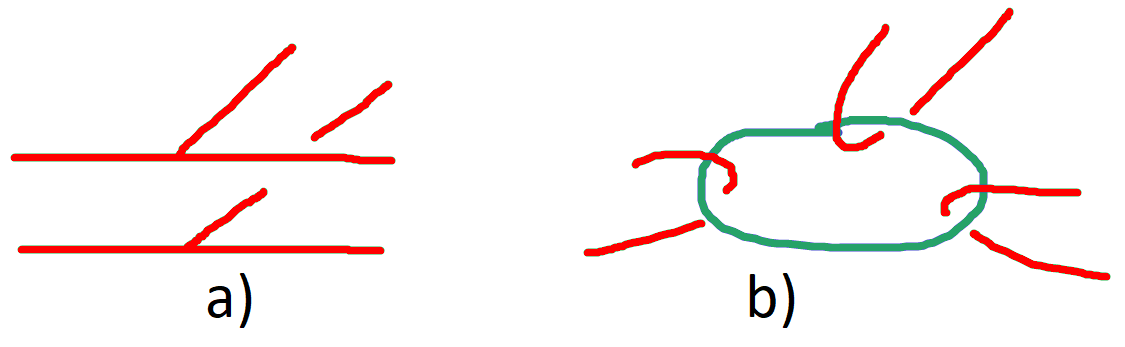}
    \caption{Structures like in a) can be decomposed into a ring of opposite type (e.g. dual) that is braided like in b) with loops. The correctness of this construction is shown by enumerating the supported correlation surfaces.}
    \label{fig:decomp}
\end{figure}

Our goal is to deconstruct arbitrary (e.g. Fig.~\ref{fig:decomp}), but valid, braided representations into loops - elementary building blocks compatible with LS. To this end, we introduce, similarly to Sec.~\ref{sec:lsb}, a \texttt{Reduced Braids Instruction Set (RBIS)}, which consists of the braiding relation and diagrammatic rewrites which leave the computation unchanged (effectively, identity operations):
\begin{enumerate}
    \item \emph{braid} strings of the same type (the identity logical operation) or of opposite type (the logical CNOT); 
    \item \emph{introduce} trivial loops;
    \item \emph{remove} trivial loops (e.g. in Fig.~\ref{fig:r2}, the loop in the middle and the strand that connects the two junctions );
    \item \emph{bridge} (e.g. in Fig.~\ref{fig:r2}b, the primal loops that are braided with the large dual);
    \item \emph{unbridge} (e.g. in Fig.~\ref{fig:r2}c, the dual strands that look like a moustache).
\end{enumerate}

\begin{figure}[!h]
    \centering
    \includegraphics[width=\columnwidth]{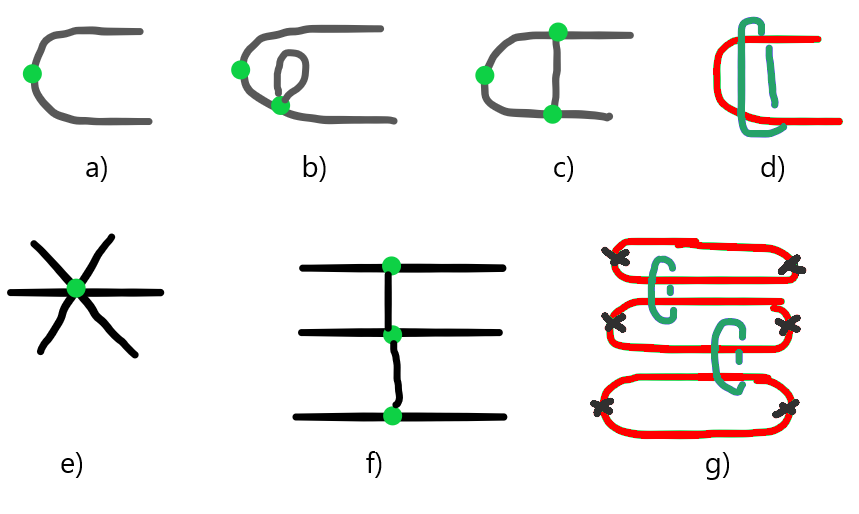}
    \caption{a) GHZ state with a green spider; b) another green spider connected to a loop can be trivially introduced; c) the loop is decomposed such that the two legs are connected; d) the braiding equivalent of the previous ZX diagram; e) a green spider with multiple legs; f) pairing the legs of the spider and decomposing the green spider; g) one the multiple possible braiding equivalents of the previous ZX diagram includes two loops braided with three logical-qubit-loops.}
    \label{fig:ring2}
\end{figure}

\begin{proposition}
    Unbridging does not affect the correctness of the diagrams.
\end{proposition}

Bridges are manifestations of the underlying CNOTs used in multibody measurement circuits. Consequently, unbridging can be used at each strand junction of a braided diagram. Trivial rings do not need to be included: these can be contracted through the ZX calculus (Fig.~\ref{fig:ring2}).

\subsection{The Raussendorf rule: The Instruction Set of Braids}
\label{sec:raussendorf}

\begin{figure}[!h]
    \centering
    \includegraphics[width=\columnwidth]{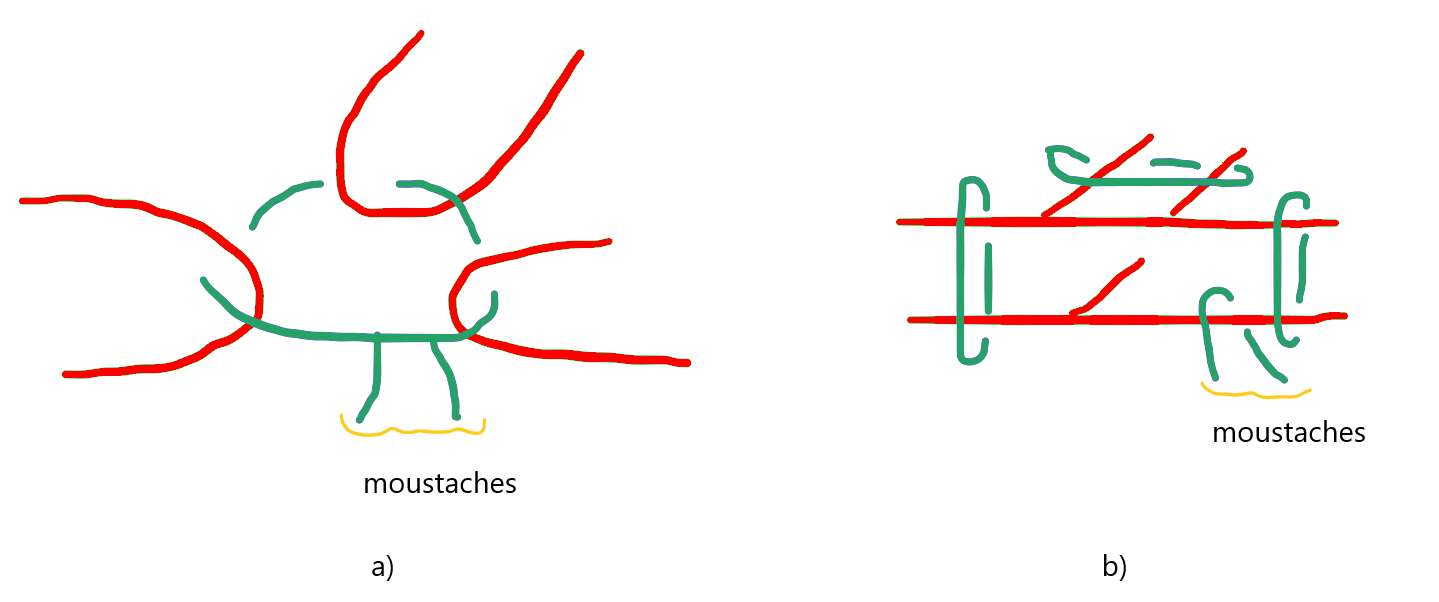}
    \caption{The compression rule as presented by Raussendorf: a) multiple loops braided with a loop of opposite type and the opposite type loop has some moustaches; b) the result of the rule is that the primals are bridged, the moustaches are unbridged and trivial dual loops are introduced.}
    \label{fig:r1}
\end{figure}

\begin{figure}
    \centering
    \includegraphics[width=\columnwidth]{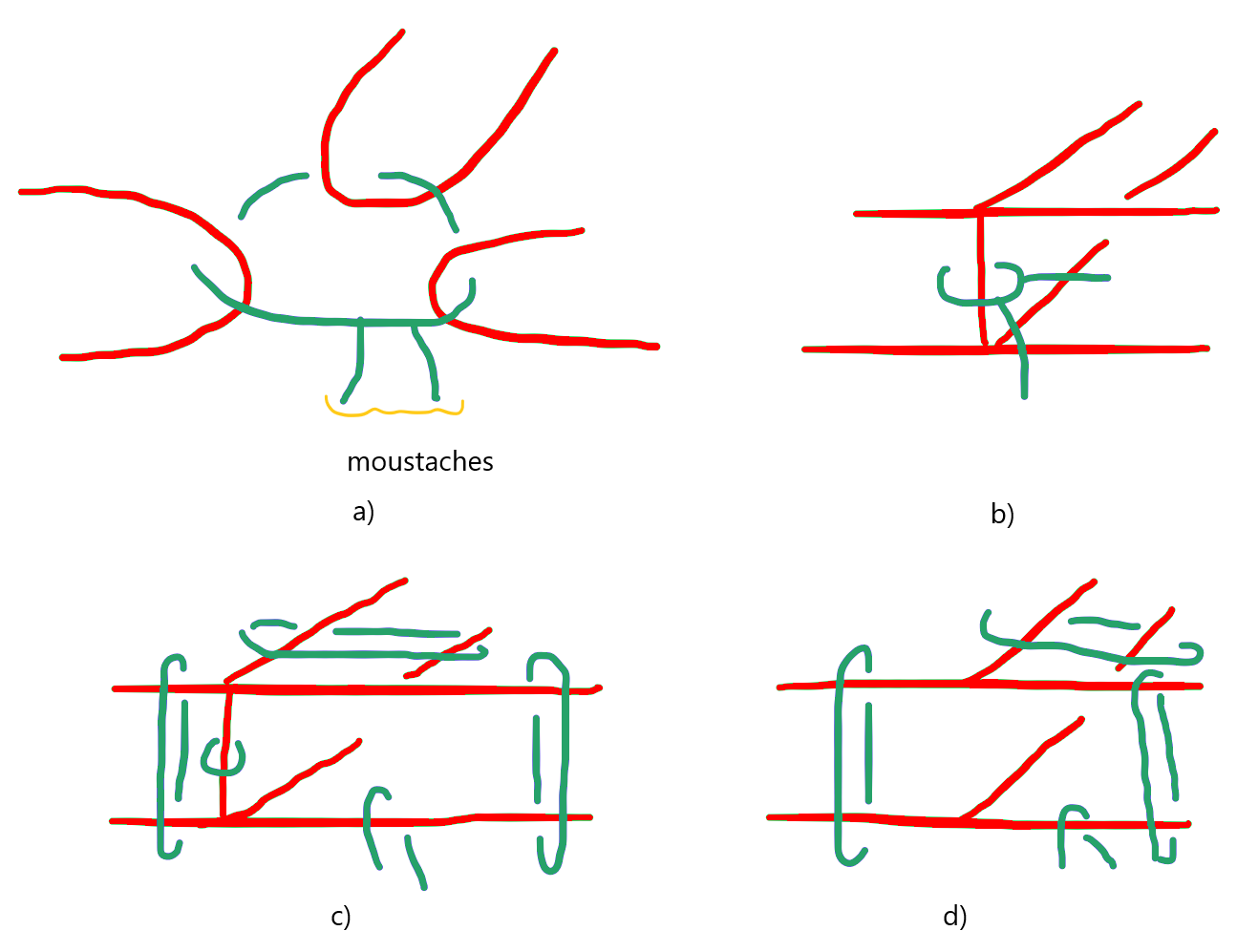}
    \caption{Step-by-step derivation of the Raussendorf rule: a) multiple loops braided with a loop of opposite type and the opposite type loop has some moustaches; b) because the blue loop correlates the green tubes, these can be bridged; c) trivial blue loops are introduced and the moustaches can be unbridged from the central blue loop; d) the central loop and the braided strand can be removed because their role in terms of correlation surfaces has been taken over by the (previously trivial) blue loops.}
    \label{fig:r2}
\end{figure}

The Raussendorf rule (Fig.~\ref{fig:r1}) encompasses all known operations about how correlation surfaces can be constructed and transformed from one into another (Fig.~\ref{fig:ring1}). In Fig.~\ref{fig:r2} we derive step-by-step the Raussendorf rule.

In terms of braided structures the Raussendorf rule can be show using the steps from Fig.~\ref{fig:r2}.

\begin{figure*}
    \centering
    \includegraphics[width=0.6\textwidth]{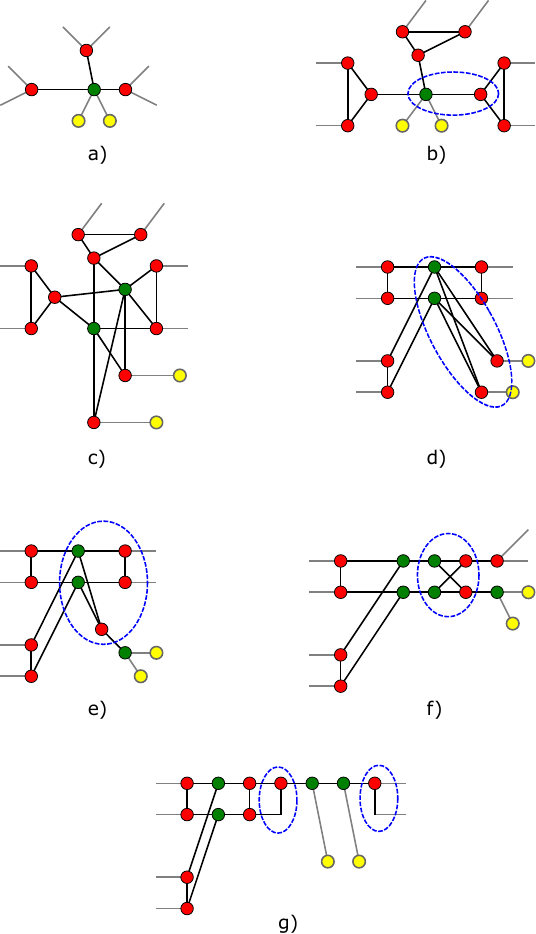}
    \caption{The steps of obtaining the Raussendorf rule with the ZX calculus: a) the original ZX diagram; b) trivial loops are introduced; c) the result of applying the bialgebra ZX rule on the spiders previously encircled blue; d) reorganising the diagram by moving the top loop to the bottom; e) the result of applying the bialgebra ZX rule on the spiders previously encircled blue; f) splitting green spiders and merging red spiders; g) the result of applying the bialgebra ZX rule on the spiders previously encircled blue. After the last step there are red encircled spiders, which represent the fact that the GHZ-like-loop logical qubit is uncomputed to a single wire. This wire is braided-CNOT with the moustaches, and then the wire is returned again to a GHZ-like-loop.}
    \label{fig:rauss}
\end{figure*}

In Fig.~\ref{fig:rauss} we used the ZX calculus and rely heavily on the bialgebra rule to perform the individual steps. We will not show the generality of the transformations formally: we will show it only for three loops. The diagrams are too complicated when more loops are involved. We have chosen to arrange the ZX diagram to look very similar to the braided structure.

\section{Discussion and Conclusion}

This work provides the first step towards a unified formal language for surface-code computation. We introduced two reduced instruction sets, \texttt{LSIS} for lattice surgery computations, and \texttt{RBIS} for braided computations, and showed how to translated between these instructions.

We have shown that arbitrary braided structure can be easily decomposed into loops of two types and that the resulting structures have the form of multibody measurements. These measurements can be translated into LS operations. From the other direction, LS operations can be easily transformed into simple braided loops. The entire process is intermediated by the ZX calculus.

In order to show that braided structures can be decomposed into simple loops, we showed that the Raussendorf rule is in fact a compressed representation of \texttt{RBIS}. The Raussendorf rule includes bridging and unbridging at the same time, as well as the introduction and removal of trivial loops.


\section*{Acknowledgements}

I am grateful for the discussions with Austin Fowler and Simon Devitt. I developed most of these ideas and translations, and large parts of the manuscript, after the long discussions I had  with Austin about compilation and decoding during my enjoyable Los Angeles 2020 pandemic stay. This research was developed in part with two Google Faculty Awards in 2019 and 2020 and a Fulbright scholarship in 2020. This research was enabled in part with funding from the Defense Advanced Research Projects Agency [under the Quantum Benchmarking (QB) program under award no. HR00112230007 and HR001121S0026 contracts], and was supported by the QuantERA grant EQUIP through the Academy of Finland, decision number 352188. The views, opinions and/or findings expressed are those of the author(s) and should not be interpreted as representing the official views or policies of the Department of Defense or the U.S. Government.

\section{Appendix}

\begin{figure}[h!]
    \centering
    \includegraphics[width=\columnwidth]{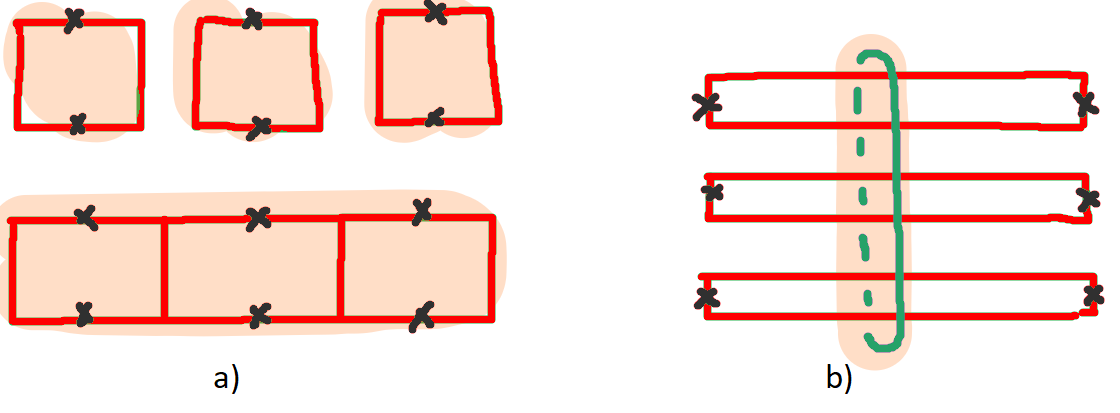}
    \caption{Multibody measurements with braids while leaving the computation unchanged: a) bridging allows the joint measurement of sheets; b) the dual ring allows the joint measurement of tubes.}
    \label{fig:xxzzloops}
\end{figure}

We discussed how to bridge and unbridge braided geometries. In Fig.~\ref{fig:unbridge} we illustrated how in some unbridged interpretations of braids, the sheets are similar to products of logical operators. Measuring for example the large sheet in Fig.~\ref{fig:unbridge}b) is equivalent to measuring the pair of the sheet operators from Fig.~\ref{fig:unbridge}c).

It is possible to use a ring to surround multiple loops of opposite type. In order to keep the discussion consistent, in Fig.~\ref{fig:xxzzloops} we use primals to illustrate how to implement XX- and ZZ- measurements. 

Therefore, in braided geometries, one introduces a tube correlation between two loops in order to measure sheet operators tuples(i.e. for primals Z-operators), and introduces support for a sheet in order to measure tube operator tuples (i.e. for primals X-operators).

\balance

\bibliography{__main}

\end{document}